# The Role of Nonequilibrium LO Phonons, Pauli Exclusion, and Intervalley Pathways on the Relaxation of Hot Carriers in InGaAs Multi-Quantum-Well Structures

Yongjie Zou[1*], Hamidreza Esmaielpour[2], Daniel Suchet[2,3], Jean-François Guillemoles[2,3], Stephen M. Goodnick[1]

[1]School of Electrical, Computer and Energy Engineering, Arizona State University, Tempe, AZ 85281, USA

[2]CNRS-Institut Photovoltaique d'Ile de France (IPVF), UMR IPVF 9006, 91120 Palaiseau, France

[3]CNRS-Ecole Polytechnique, UMR IPVF 9006, 91120 Palaiseau, France.

[*]Email: yongjie.zou@protonmail.com

## Abstract

Under continuous-wave laser excitation in an InGaAs multi-quantum-well (MQW) structure, the carrier temperature extracted from photoluminescence rises faster for 405 nm excitation compared with 980 nm excitation, as the injected carrier density increases. Ensemble Monte Carlo simulation of the carrier dynamics in the MQW system shows that this carrier temperature rise is dominated by nonequilibrium LO phonon effects, with the Pauli exclusion having a significant effect at high carrier densities. Further, we find a significant fraction of carriers reside in the satellite L-valleys for 405 nm excitation due to strong intervalley transfer, leading to a cooler steady-state electron temperature in the central valley compared with the case when intervalley transfer is excluded from the model. Good agreement between experiment and simulation has been shown, and detailed analysis has been presented. This study expands our knowledge of the dynamics of the hot carrier population in semiconductors, which can be applied to further limit energy loss in solar cells.

## I. INTRODUCTION

In a conventional solar cell, the excess kinetic energy of photoexcited electrons and holes is lost through energy loss processes (involving phonons) and they relax to their respective bandedges, where they are extracted through Ohmic contacts. The loss of the excess energy of photons relative to the material bandgap, i.e. thermalization loss, is one of the fundamental losses that lead to a maximum theoretical conversion efficiency of 33% and 41% for a single-junction solar cell under 1 sun and maximum-concentration blackbody solar radiation, respectively [1]. Multijunction solar cells with multiple bandgaps reduce thermalization loss by absorbing different portions of the solar spectrum in different-bandgap materials, leading to efficiencies in excess of 45% under concentration [2]. The ultimate goal is to harness as much energy from the photogenerated carriers as possible by avoiding thermalization loss. Hot carrier solar cells are one approach among many advanced concept proposals to avoid thermalization loss by extracting carriers not at the band edges, but through contacts made to the electrons and holes at higher energies (selective energy contacts), that extract hot carriers from the device before they have time to lose their excess energy [3]–[5]. There have been numerous investigations using zincblende materials as absorbers for hot carrier solar cells, as the advancement of epitaxial growth technology for these materials allows the design of layer structures avoring enhanced hot carrier effects. In these polar materials, the Fröhlich interaction between the carriers and longitudinal optical (LO) phonons is the dominant carrier-phonon scattering mechanism. An LO phonon can typically decay into two acoustic phonons (Klemens [6]) or into one optical phonon and one acoustic phonon (Ridley and Gupta [7] or Vallée and Bogani [8]). These three-phonon interactions are usually much slower than the rate of LO phonon emission by carriers. Therefore, at high injected carrier densities, the population of LO phonons due to emission from hot electrons can be driven strongly out of equilibrium compared to the equilibrium Bose-Einstein distribution, leading to the so-called "phonon bottleneck". The nonequilibrium LO phonon population can be re-absorbed by the carriers, maintaining the excess energy in the system by keeping the carriers hot, i.e. with an energy much higher than the equilibrium temperature of the lattice [9]–[11].

The Pauli exclusion principle may also play a role in slowing carrier cooling at high carrier densities, where filling of low energy states can reduce the number of final states that high-energy carriers can relax into. Investigations of Pauli blocking effects are typically reported for transient properties, such as the ultrafast optical response[12]–[14]. Esmaielpour *et al.* discussed this effect on the accuracy of temperature extraction from the slope of photoluminescence spectra.[15]

Moreover, electrons with energies higher than the satellite valleys have a high probability of transferring to the satellite valleys due to the high density of states compared to the central valley, before transferring back to the central valley. The transfer of photoexcited carriers into satellite valleys in InAs based materials is the basis for a recently proposed alternate approach to hot carrier solar cells, so-called valley photovoltaics.[16]–[19] However, whether these intervalley pathways contributes to a higher steady-state electron temperature in the central valley has not been fully discussed in literature.

Reduced cooling of photogenerated carriers have been primarily observed in quantum well structures.[11], [20], where a review of the early work in the field is given in [21]. More recently, researchers have demonstrated extraction of hot carriers from quantum well absorbers[22], [23], and a quantum well hot-carrier solar cell has reached 11% conversion efficiency[24]. Research in this area has also been recently reviewed[3].

In the present work, we first experimentally extract the carrier temperature in a type-I band alignment $In_{0.53}Ga_{0.47}As/In_{0.8}Ga_{0.2}As_{0.44}P_{0.56}$ multi-quantum-well (MQW) structure from photoluminescence characterization illuminated by continuous-wave (CW) lasers at 405 nm (above the L valley) and at 980 nm (below the L valley). We then use the experimental data as a baseline for theoretical analysis with the ensemble Monte Carlo (EMC) method. The EMC method is a particle-based solution to the Boltzmann transport equation, which simulates particle motion in the reciprocal space and/or the real space, where they are assumed to undergo free-flights terminated by random scattering processes.[25], [26] By employing different combinations of fundamental mechanisms in EMC simulations, one can distinguish the effects of these underlining mechanisms on the ensemble results. The EMC method has been successfully used to study hot carrier relaxation in quasi-2D systems[27], [28]. In particular, we focus on using EMC simulations to understand the effects of nonequilibrium LO phonons, Pauli exclusion, and intervalley pathways on the steady-state temperature of the electrons in the $\Gamma$ valley in comparison with experiment.

## II. EXPERIMENT

### A. Sample Preparation

The InP-lattice matched MQW structure consists of five un-doped $In_{0.53}Ga_{0.47}As$ QWs (5.5 nm) and $In_{0.80}Ga_{0.20}As_{0.44}P_{0.56}$ (10 nm) barriers. The MQW is cladded by an InP cap (30 nm) and the InP substrate to improve the accumulation of photo-generated carriers within the active region. The MQW structure is grown by molecular beam epitaxy (MBE) on an un-doped InP substrate. The schematic of the sample is shown in Fig. 1(a).

## B. Photoluminescence Characterization

Hot carrier properties of a In$_{0.53}$Ga$_{0.47}$As/In$_{0.80}$Ga$_{0.20}$As$_{0.44}$P$_{0.56}$ MQW structure [Fig. 1(a)] are determined via continuous-wave photoluminescence (PL) spectroscopy. The sample is excited by two laser lines, 405 nm and 980 nm, which excite electrons from the valence band to energy levels above and below the L-valley, respectively, in the conduction band of the InGaAs QW. The temperature of the lattice is fixed at 300 K. The PL emission by the sample is detected by a hyperspectral luminescence imager, which can create spectrally (2 nm) and spatially resolved PL maps. The PL spectra emitted by the InGaAs MQW under various excitation powers at the center of the concentrated laser spot are shown in Fig. 1(b).

The thermodynamic properties of emitting particles are determined by fitting the whole PL spectrum with the generalized Planck's law, as described by[29], [30]:

$$I_{PL}(E) = \frac{2\pi A(E)(E)^2}{h^3 c^2}\left[exp\left(\frac{E - \Delta\mu}{k_B T}\right) - 1\right]^{-1}, \quad (1)$$

where $I_{PL}$ is the PL intensity, $A(E)$ energy-dependent absorptivity, $c$ the speed of light, $h$ the Planck's constant, $k_B$ the Boltzmann constant, and $\Delta\mu$ represents the difference of the quasi-Fermi levels of the electrons and holes ($\Delta\mu = \mu_c - \mu_v$) under photo-excitation conditions. The details of the full spectrum fit are presented elsewhere[24], [31]. The results of the full spectrum fit are shown by the black lines in Fig. 1(b). The energies of the optical transitions are labeled on the PL spectra. From the full spectrum fit, it is observed that there are three dominant optical transitions in the sample (one excitonic and two band-to-band transitions). Fig. 1(c) indicates the power-dependent behavior of the energy values of the optical transitions. It is seen that the energy values of the optical transitions do not change significantly with power, which is consistent with the type-I band alignment of the QW structure.

The results of hot carrier temperature difference ($\Delta T$: the temperature difference between hot carriers and the lattice) determined from the full spectrum fit are plotted in Fig. 1(d). The 2D steady-state carrier density is determined by multiplying a carrier lifetime of 3 ns by the absorbed photon flux for the two laser lines, which is determined by dividing the amount of absorbed power within the active region by the energy of each photon, and the absorbed power in the active region is determined by the transfer matrix method[32], [33].

Figure 1(d) indicates that by increasing the steady-state carrier density, the temperature of hot carriers becomes higher under both excitation wavelengths. However, it is seen that at higher steady-state carrier densities, the increase in the temperature of hot carriers for the incident wavelength at 405 nm is greater than that for the excitation at 980 nm.

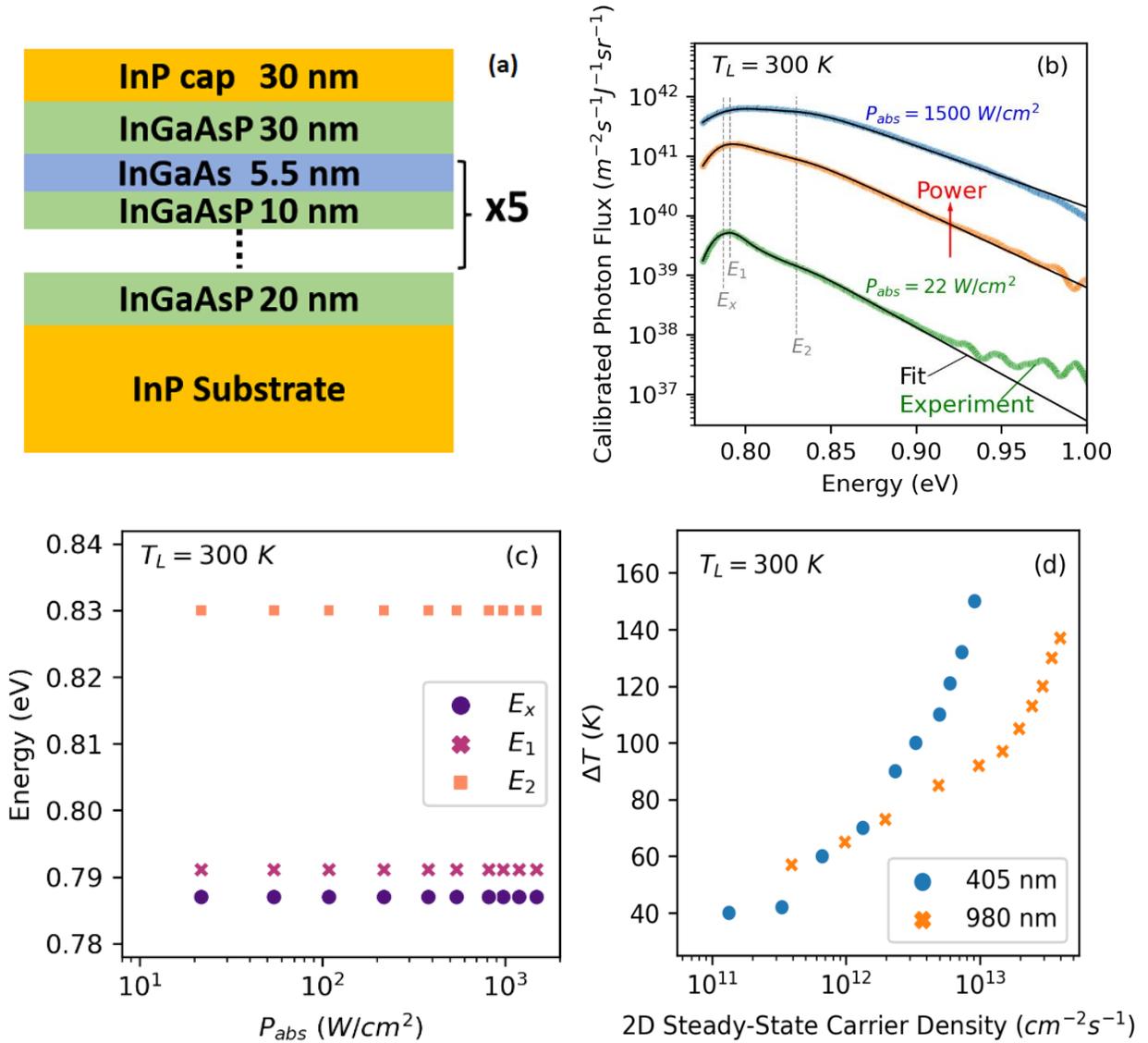

Fig. 1. Photoluminescence (PL) characterization of the multi-quantum well (MQW) structure. (a) Schematic of the MQW structure (not to scale). (b) PL spectra of the InGaAs MQW structure at 300 K under various excitation powers of the 405 nm laser. The black lines indicate results of the full spectrum fit. The energy positions of optical transitions in the QW structure are labeled by $E_x$, $E_1$, $E_2$. The $E_x$ is due to excitons. (c) The optical transition energies as a function of absorbed power density. (d) The hot carrier temperature difference $\Delta T$ versus the 2D steady-state carrier density at 300 K under 980 nm and 405 nm excitation wavelengths.

## III. THEORETICAL SIMULATION

### A. Electronic Band Structure

To simulate carrier dynamics in this MQW structure using the ensemble Monte Carlo (EMC) method, we calculate the electronic band structure by solving the single band effective mass equation in the growth direction, while motion in the plane of the quantum wells is unconfined, described by an in-plane effective mass.

Table 1. Material parameters at 300 K, used for the band structure calculation. They are based on interpolation from Ref.[34], band offset studies[40], and adjustment for quantization effects[35]–[39] according to measured photoluminescent energy peaks in this work. ∥ and ⊥ denote directions parallel and perpendicular to the hetero interfaces, respectively.

|  | $In_{0.53}Ga_{0.47}As$ | $In_{0.80}Ga_{0.20}As_{0.44}P_{0.56}$ |
|---|---|---|
| $E_\Gamma$ (eV) | 0.73 | 0.85 |
| $E_L$ (eV) | 1.28 | 1.46 |
| $E_{HH}$ (eV) | 0.00 | -0.20 |
| $m_{\Gamma,\perp}/m_0$ | 0.061 | 0.062 |
| $m_{L,\perp}/m_0$ | 0.438 | 0.447 |
| $m_{HH,\perp}/m_0$ | 0.712 | 0.720 |
| $m_{\Gamma,\parallel}/m_0$ | 0.056 | 0.062 |
| $m_{L,\parallel}/m_0$ | 0.393 | 0.447 |
| $m_{HH,\parallel}/m_0$ | 0.674 | 0.720 |

The $\Gamma$ valley and the L valley of the conduction band and the heavy hole (HH) band are included in the EMC simulations, while other satellite valleys or bands are excluded from the simulation domain for simplicity. The well material in the structure, $In_{0.53}Ga_{0.47}As$ is a direct bandgap material, i.e. the bottom of conduction band aligns with the top of the valence band at the $\Gamma$ symmetry point. Since the HH band has a large density of states compared to other hole bands, most radiative recombination occurs between HH and the $\Gamma$ valley of the conduction band. Carriers with energies above the bottom of the L valley of the conduction band have a high probability of being scattered to the L valley, due to the effective mass in the L valley (considering all equivalent valleys) being much larger than that in the $\Gamma$ valley. Table 1 lists the material parameters used for the MQW band structure calculation. These parameters are approximated by interpolation from the band parameters in the literature[34], and then the normal effective mass $m_{\Gamma,\perp}$ for the wells are adjusted so that the ground state energy in the wells is close to the $E_1$ PL peak, the parallel effective mass $m_{\Gamma,\parallel}$ for the wells are adjusted for quantization effects[35]–[39], and with a lack of knowledge the other effective masses for the wells are modified proportional to the changes for their $\Gamma$ masses. These adjustments can be attributed to non-parabolicity of the bands and the wells' wavefunction penetration into the barriers[35]–[39], while considering there could be some inaccuracy in the interpolation as well. Fig. 2 shows the calculated subband energy levels and their corresponding wavefunctions. Ten subbands for each valley are included in our EMC simulations corresponding to two groups of five, with each of the weakly coupled wells contribute primarily to two subbands. Excitonic effects are ignored in this work, as they are assumed small at 300 K and above. The main band-to-band optical transition $\Gamma$1-HH1 is calculated to be 0.794 eV, which is close to the measured PL peak $E_1 \approx 0.79$ eV. The $\Gamma$1-HH2 is calculated to be 0.828 eV, close to the measured PL peak $E_2 \approx 0.83$ eV. But the $\Gamma$1-HH2 transition is very weak due to symmetry and should not play any significant role.

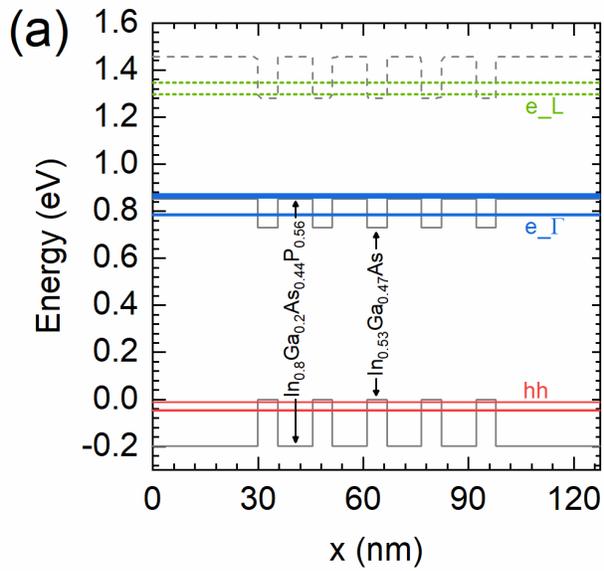
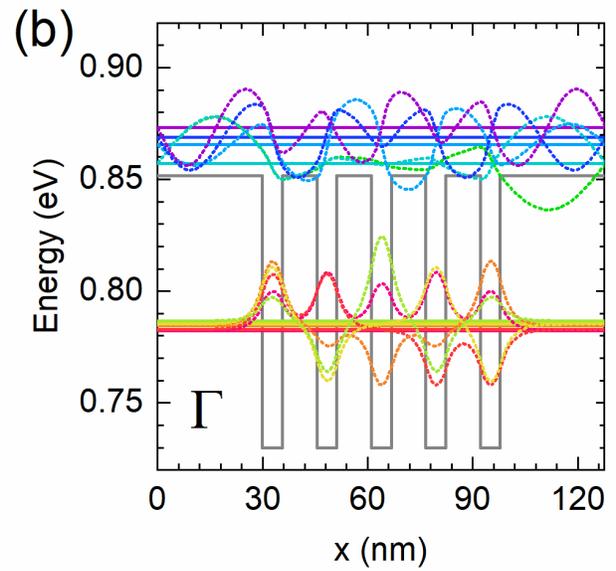
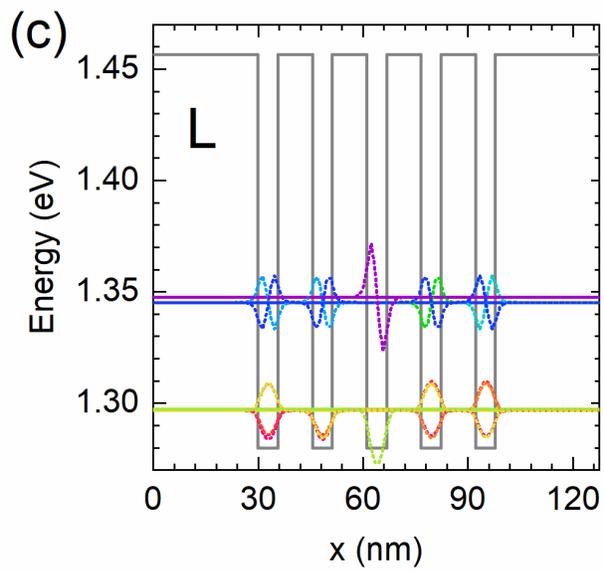
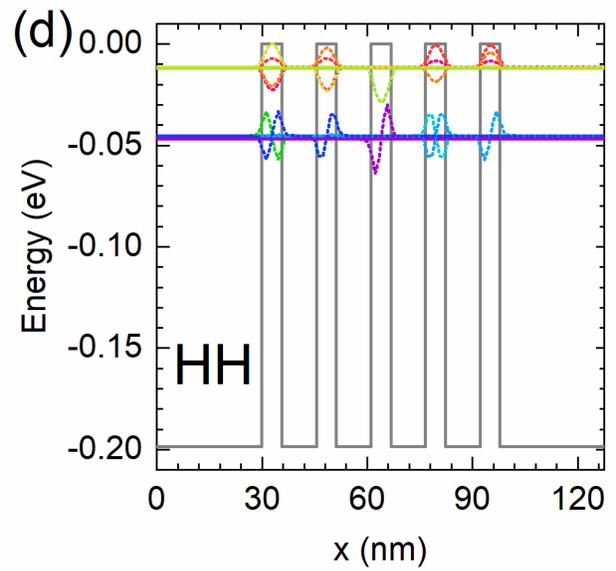

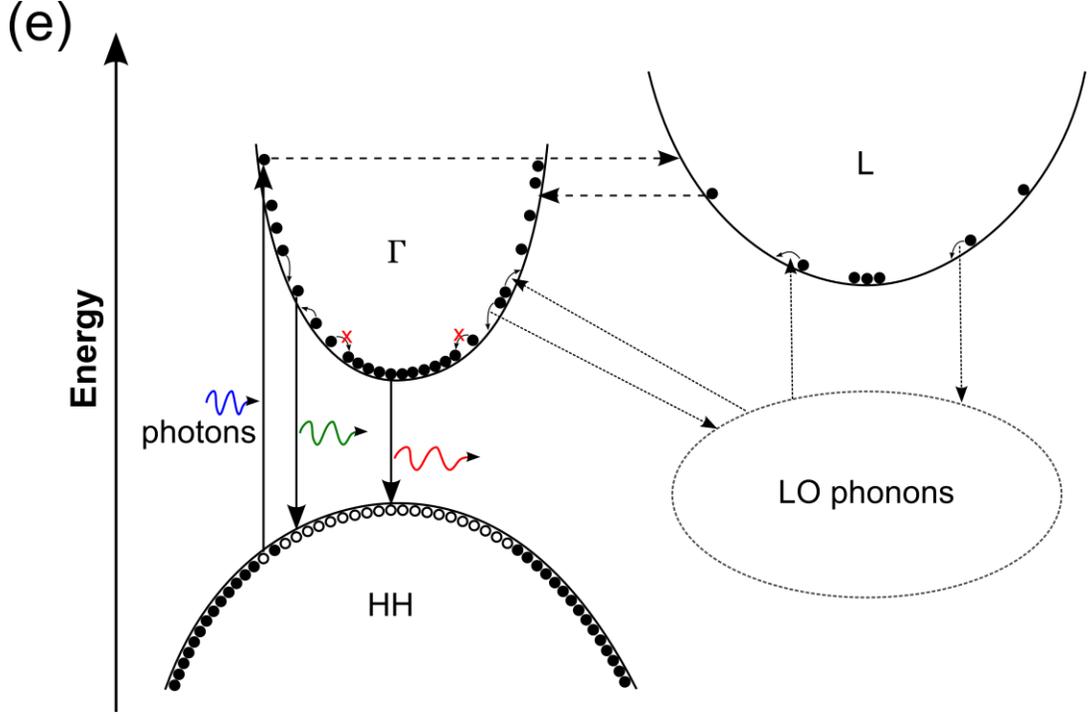

Fig. 2. The ensemble Monte Carlo simulation space and mechanisms. (a) The calculated energy levels for the MQW. (b), (c), and (d) show the energy levels (horizontal solid lines) and wavefunctions (dotted curves) of the subbands perpendicular to the layer interfaces for the electrons ($\Gamma$ and L) and heavy holes (HH). (e) The optical generation and recombination, electron-LO phonon interaction, and intervalley scattering processes included in the EMC simulations.

## B. Ensemble Monte Carlo Simulations

Following the experiment, the EMC simulations assume the lattice temperature is 300 K, and consider two continuous excitation wavelengths, 405 nm and 980 nm, which excite electrons either above or below the bottom of the L valley, respectively. Figure 2e illustrates schematically the dynamical balance between electrons, holes and optical phonons, where photons create electron-hole pairs, high energy carriers can scatter to and from the L valley from the $\Gamma$ valley, and electrons (and holes) exchange energy with the phonon bath primarily through the polar optical phonon scattering, driving the phonon system out of equilibrium.

CW optical carrier generation is modeled by adding at each timestep a fixed number of electron-hole pairs into the $\Gamma$ valley and the HH band according to each specified generation rate with a Gaussian spectral profile centered around the beam excitation energy. The generation probabilities between subbands are proportional to the oscillator strength due to wavefunction overlap. Recombination (radiative, traps, etc.) is simulated by random selection of electron-hole pairs to be removed from the $\Gamma$ valley and the HH band through a simple decay. In the commonly used relaxation time approximation, the reduction of carrier density due to recombination at each timestep is given by,

$$\varDelta n = \frac{-\varDelta t}{\tau_r}(n - n_0),$$
(2)

where $n_0$ is the equilibrium carrier density, $n$ is the carrier density before the recombination, $\Delta t$ is the timestep, $\tau_r$ is the recombination lifetime, and $\Delta n$ is the change in carrier density due to recombination. As shown in the experiment, the hot carrier temperature strongly depends on the steady-state carrier density. Therefore, we tried to achieve similar steady-state carrier densities in the simulations compared to the experiment. We use a small $\tau_r = 0.1$ ns to allow the simulations to reach steady state within our computation capacity, and then adjust the generation rates accordingly for different target steady-state carrier densities. This time constant is still much larger than the LO phonon emission time, and therefore is not expected to affect the dynamics. However, the assumption for generate rates may affect the simulation results, and the impact is planned to be investigated in another work that requires more computation power.

The scattering mechanisms for the EMC simulations include carrier-carrier (electron-electron, hole-hole and electron-hole) scattering, carrier-longitudinal optical (LO) phonon scattering, and intervalley scattering between $\Gamma$ and L[26] (later we will also show a case where the intervalley scattering is switched off). These processes are illustrated in Fig. 2(e).

The main energy loss mechanism for carriers in III-V materials is the emission of LO phonons[41]. Due to their low group velocity, the emitted LO phonons are assumed to remain within the excitation volume, where they are either re-absorbed by the carriers, or undergo anharmonic decay into acoustic phonons, which are assumed to rapidly propagate out of the system. The decay of LO phonons into acoustic phonons is modeled by the rate equation at each timestep as [27]

$$\Delta N_q = \frac{-\Delta t}{\tau_{LO}}(N_q - N_{q0}),$$
(3)

where $N_{q0}$ is the equilibrium LO phonon population, $N_q$ is the LO phonon population before the decay, $\Delta t$ is the timestep, $\tau_{LO}$ is the LO phonon lifetime, and $\Delta N_q$ is the change in LO population due to the decay. This form is identical to the observed decay observed in transient Raman studies measuring the LO phonon lifetime [42-46].

The measured LO phonon decay lifetimes for GaAs and InAs at 300 K are reported to be about 3.5 ps[42]–[45] and 1.8 ps[46], respectively. We use $\tau_{LO} = 2.5 ps$ for In$_{0.53}$Ga$_{0.47}$As as a linear interpolation between the two values. To understand the role of the LO phonon lifetime on carrier heating, we also consider the case when nonequilibrium LO phonons (NELO) are neglected ($N_q \equiv N_{q0}$), and the case when the LO phonon lifetime is much larger, $\tau_{LO} = 25 ps$. The reported deformation potentials of scattering between the $\Gamma$ and L valleys, $D_{\Gamma,L}$ in GaAs and in InAs averaged around 3.96 eV/Å and 2.29 eV/Å, respectively[47]–[50]. Assuming a linear interpolation, we use $D_{\Gamma,L} = 3.1$ eV/Å for In$_{0.53}$Ga$_{0.47}$As.

### C. Electron Population in the $\Gamma$ and L Valleys

The 980 nm excitation generates electrons below the bottom of the L valley, allowing only a small fraction of the $\Gamma$ electrons to scatter into the L valley. In this case, an extremely high density of $\Gamma$ electrons are needed to appreciably populate the L valley. Fig. 3 shows the EMC results of electron densities in the $\Gamma$ and L valleys as a function of time, with the 2D steady-state densities in the $\Gamma$ valley, $n_{ss,\Gamma}^{2D} \cong 4 \times 10^{13} cm^{-2}$. With 980 nm CW excitation and no NELO, only roughly 0.25% of the electrons appear in the L valley at steady state. The 405 nm excitation lies well above the L valley bottom, the electrons generated in $\Gamma$ have very high probabilities of being scattered into the L valley. In fact, at the early stages of the 405 nm simulations, more electrons appear in the L valley compared to in the $\Gamma$ valley. But over time, more and more $\Gamma$ electrons lose energy by emitting multiple phonons, and their chance of being scattered into the L valley becomes smaller; therefore, more electrons build up in the $\Gamma$ valley. At steady state, the exchange between $\Gamma$ and L keep a larger population in $\Gamma$ than in L. The counterpart cases with NELO are shown in Fig. 3(b). The absorption of NELO by electrons increases their occupation in higher energy states, leading to more electrons to be scattered into the L valley. For the 980 nm case, L valley population is raised more than two folds, compared to when there is no NELO accounted. The L valley population also increases in the 405 nm case for the same reason.

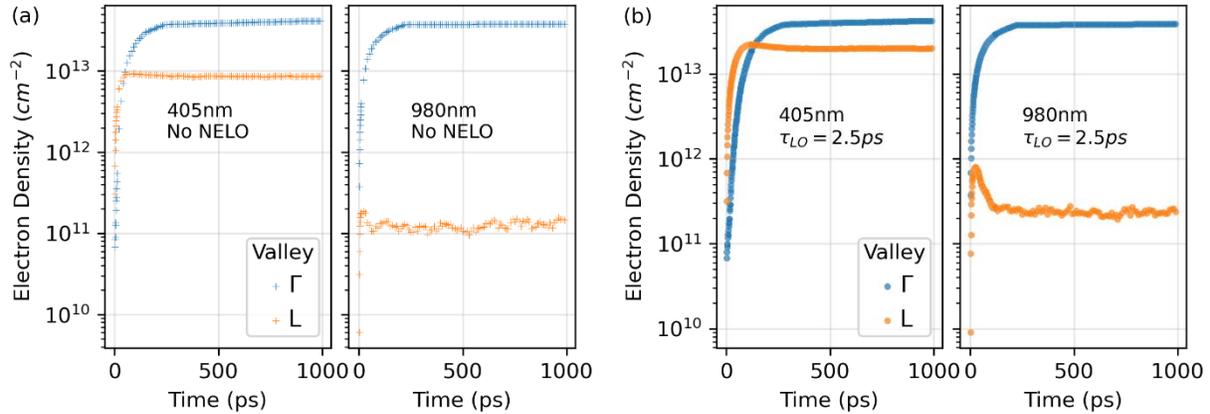

Fig. 3. EMC time evolution of electron densities in the $\Gamma$ and L valleys under 405 nm and 980 nm CW excitation. (a) Results without nonequilibrium LO phonons. (b) Results with nonequilibrium LO phonons.

### D. Electron Distribution and Temperature Fitting

As the EMC simulations stochastically solve the Boltzmann transport equation, we obtain the carrier distribution as a function of time. As the carriers equilibrate with each other, they form a Fermi-Dirac distribution, which, at high temperature can be approximated as a Maxwell-Boltzmann distribution for the high energy tail. In this work, we focus on the carrier distribution in the $\Gamma$ valley, which can be a source from which a hot-carrier current is extracted. Fig. 4(a) plots the electron distribution in the lowest ten subbands of the $\Gamma$ valley after the simulation has reached steady state. The electrons have essentially the same distribution in the lowest five subbands in the MQW. Compared to the lowest five subbands (below the barriers), the higher five subbands (above the barriers) are less populated as expected and have similar electron distribution slopes. This closeness of the distribution slopes is probably caused by the fast electron-electron interaction, which drives the electrons towards a common thermal equilibrium. Since this is typically observed across our simulated parameter space, we focus on analyzing carriers in the first $\Gamma$ subband ($\Gamma 1$) from here. The $\Gamma 1$ distribution increases as $n_{ss,\Gamma}^{2D}$ increases. Significant state filling happens when $n_{ss,\Gamma}^{2D}$ is high, above $4.5 \times 10^{12} cm^{-2}$ for the case shown in Fig. 4(b). Note the arrow in Fig. 4(b) crosses the distributions at their respective Fermi energies. We will discuss the state filling effect via Pauli exclusion later.

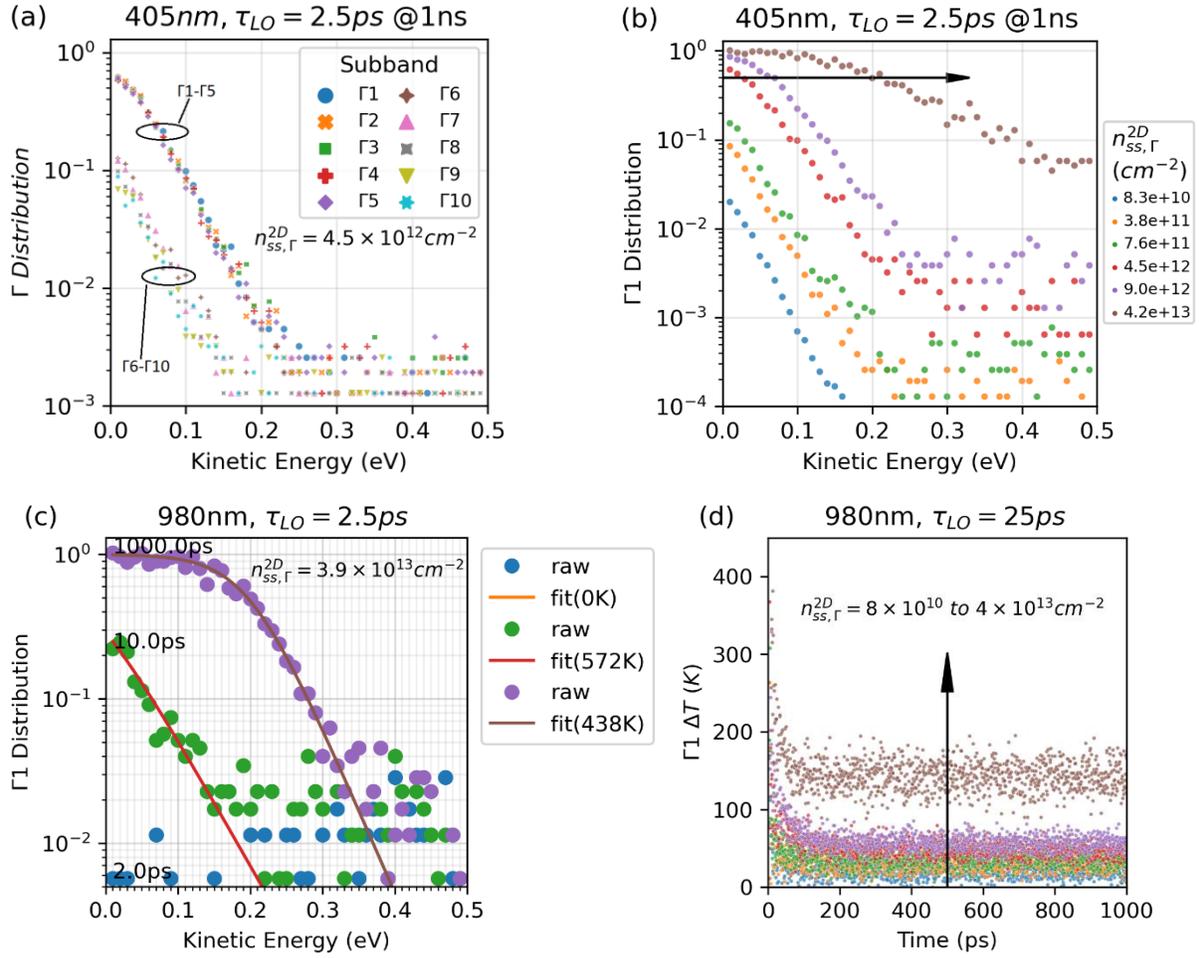

Fig. 4. Electron distribution and temperature fitting. (a) Simulated electron distribution in the $\Gamma$ valley at 1 ns for the lowest ten subbands at $n_{ss,\Gamma}^{2D} = 4.5 \times 10^{12} cm^{-2}$. (b) Simulated electron distribution in the $\Gamma 1$ subband at various $n_{ss,\Gamma}^{2D}$. The arrow in indicates the Fermi energy increases with $n_{ss,\Gamma}^{2D}$. (c) Electron distribution in subband $\Gamma 1$ at 2 ps, 10 ps, and 1000 ps as simulated from EMC (solid cycles) and as Fermi-Dirac fits (lines). The fitted absolute temperatures are indicated in the legend. Failure to fit is default to 0 K. (d) The fitted temperature difference $\Delta T$ as a function of time, with 980 nm excitation and $\tau_{LO} = 2.5 ps$. The colors from blue, orange, green to red, purple, coffee are coded for $n_{ss,\Gamma}^{2D} = 8 \times 10^{10}, 4 \times 10^{11}, 8 \times 10^{11}, 4 \times 10^{12}, 8 \times 10^{12}, 4 \times 10^{13} cm^{-2}$.

The fast electron-electron interaction allows the electrons to thermalize to a Fermi-Dirac distribution on a time scale around 0.1 ps, and these electrons lose energy to phonons with a similar time constant. Although CW excitation keeps adding newly generated electrons to the system, our EMC results suggest the majority of the electrons still have close to Fermi-Dirac distributions during most of the simulation period. This allows us to estimate the electron temperature $T_e$ by fitting the electron distribution to the Fermi function of their kinetic energy,

$$f(E) = \frac{1}{e^{\frac{E-\mu}{k_B T_e}} + 1},$$

(4)

where $E$ is the electron kinetic energy, $\mu$, also a fitted parameter, is the Fermi energy of the electrons with respect to the bottom of the energy subband, $k_B$ is the Boltzmann constant. The electron distribution in $\Gamma 1$ with 980 nm CW excitation is plotted in Fig. 4(c). At 2 ps, most of the generated electrons are still near but below the excitation energy, which is 0.42 eV above the bottom of $\Gamma 1$. At this early stage, reliable temperature fitting is difficult to achieve, and the temperature is arbitrarily default to 0 K. As the electrons lose energy via emitting more phonons, the majority of the electrons build up close to the subband bottom, and the Fermi-Dirac temperature fitting becomes possible in these energy ranges, for the case shown in Fig. 4(c), 0–0.13 eV for 10 ps and 0–0.31 eV for 1000 ps. The slopes of the distribution near the higher end of these energy ranges become steeper as the simulation progresses, and reach more stable values as the simulation arrives at steady state. The fitted temperature is 134 K higher for 10 ps than for 1000 ps, while the subband is much more filled at the latter timestamp. Fig. 4(d) plots the fitted $\Gamma 1$ electron temperature difference as a function of simulation time. All simulations in this work reach steady state before 600 ps.

### E. Scattering Rates of $\Gamma$ electrons

Figure 5 plots the scattering rates in the $\Gamma$ valley with and without Pauli exclusion enabled. At the beginning, the $\Gamma$-L intervalley scattering rates are among the highest. As more and more electrons appear in the L valley at the early stage, the net electron transfer from $\Gamma$ to L slows down, and more $\Gamma$ electrons interact with LO phonons, the so called Polar Optical Phonon (POP) scattering, giving rise to the decrease in $\Gamma$-L rates and the increase in POP rates. The dip in the electron-electron (e-e) rate is probably due to an increasingly larger portion of the electrons are interacting with phonons at the time, both intravalley and intervalley. This kind of dips happen for all simulated injection intensities, and concur with the increase in the electron-phonon scattering rates at similar time lengths (not shown here for space concern). The e-e rate then rises as the electron density increases, and then slows down again due to screening at high electron densities; such screening effect is not seen in lower-injection cases. At steady state in cases with $n_{ss,\Gamma}^{2D} > 4 \times 10^{12} cm^{-2}$, the POP and e-e rates are lower with Pauli exclusion included in the simulations, because the low energy electrons block the higher energy electrons from being scattered into those occupied low energy states. This effect on the POP and e-e rates is not obvious when $n_{ss,\Gamma}^{2D} \leq 8 \times 10^{11} cm^{-2}$. The $\Gamma$-L rates remain essentially the same with or without Pauli exclusion, since electrons occupying low energy states do not affect the $\Gamma$-L exchange.

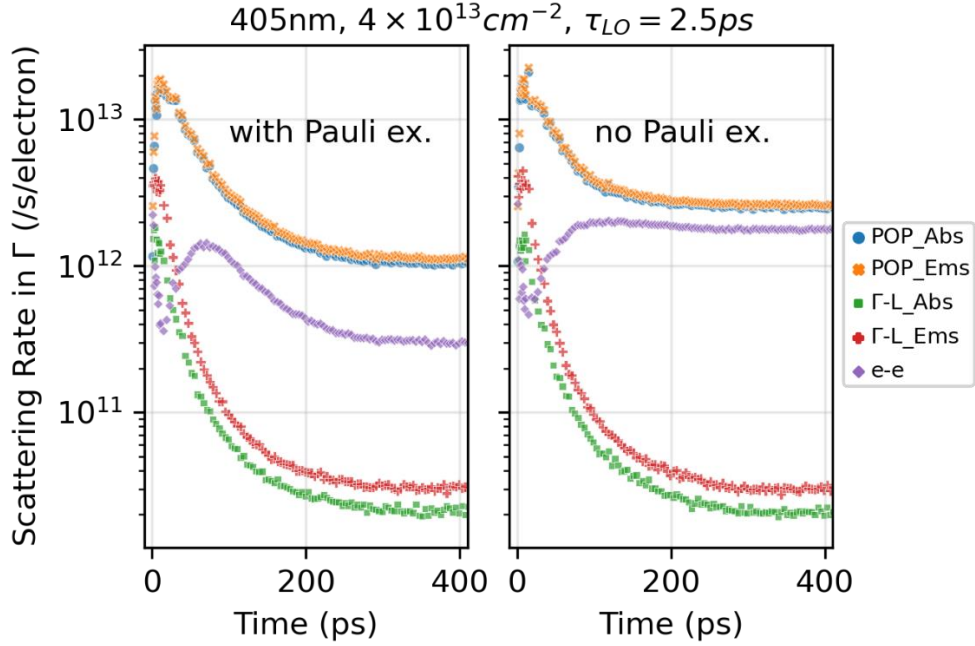

Fig. 5. Scattering rates in the $\Gamma$ valley with (left panel) and without (right panel) Pauli exclusion. Both cases have $n_{ss,\Gamma}^{2D} = 4.2 \times 10^{13} cm^{-2}$.

### F. Steady-State Temperature of $\Gamma$ Electrons

The steady-state hot electron temperature differences are calculated by averaging the fitted temperatures such as those in Fig. 4(d), over time for the last 200 ps. The results are plotted in Fig. 6 for different NELO relaxation times. Without NELO [Fig. 6(a)], the steady-state $\Gamma 1$ electron temperatures for both the 405 nm and 980 nm excitation scenarios are roughly the same as the lattice temperature, considering fitting error. However, for extremely high $n_{ss,\Gamma}^{2D}$ of about $4 \times 10^{13} cm^{-2}$, the Pauli exclusion effect becomes significant as discussed above, and the electron temperature rises above the lattice temperature. When NELOs are enabled [Figs. 6(b) and 6(c)], i.e., nonequilibrium LO phonon decay time $\tau_{LO} > 0$, more LO phonons can be absorbed by the electrons, leading to higher electron temperatures. Compared to the 980 nm cases, photo-generated electrons under 405 nm excitation have higher energies, and thus emit more LO phonons, causing a stronger NELO effect. As a result, the 405 nm cases show higher steady-state temperatures of the $\Gamma 1$ electrons. This trend agrees with experiment observation discussed above. The data from PL characterization are added to each sub-figure as symbols for comparison. The lines in Fig. 6(b) are simulated with a realistic $\tau_{LO} = 2.5\ ps$ and show the best agreement with the experiment, among the simulated scenarios.

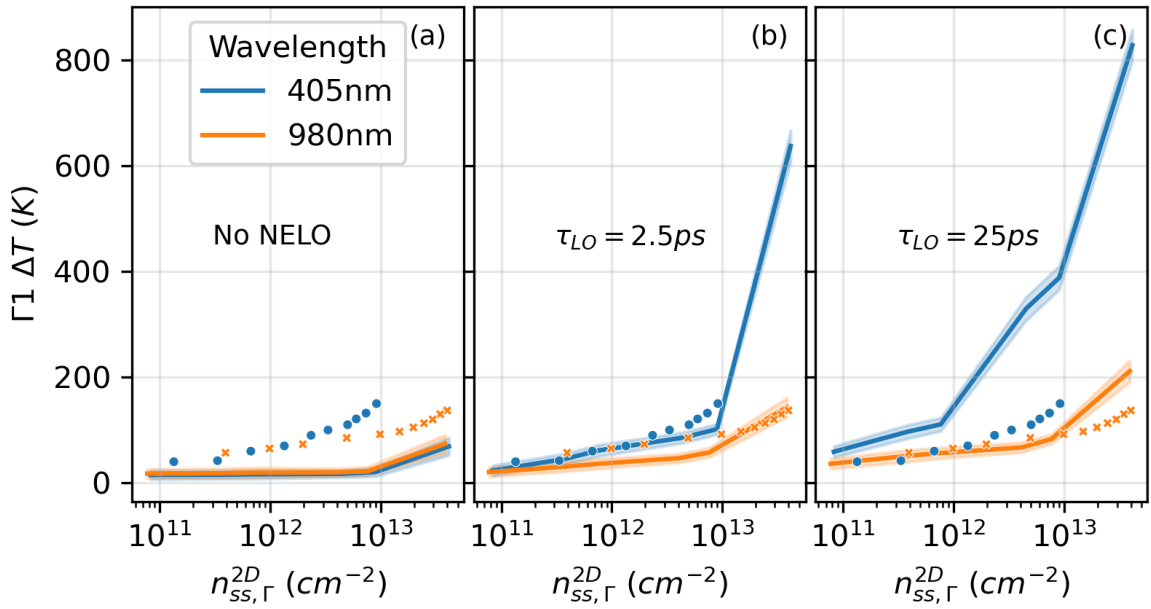
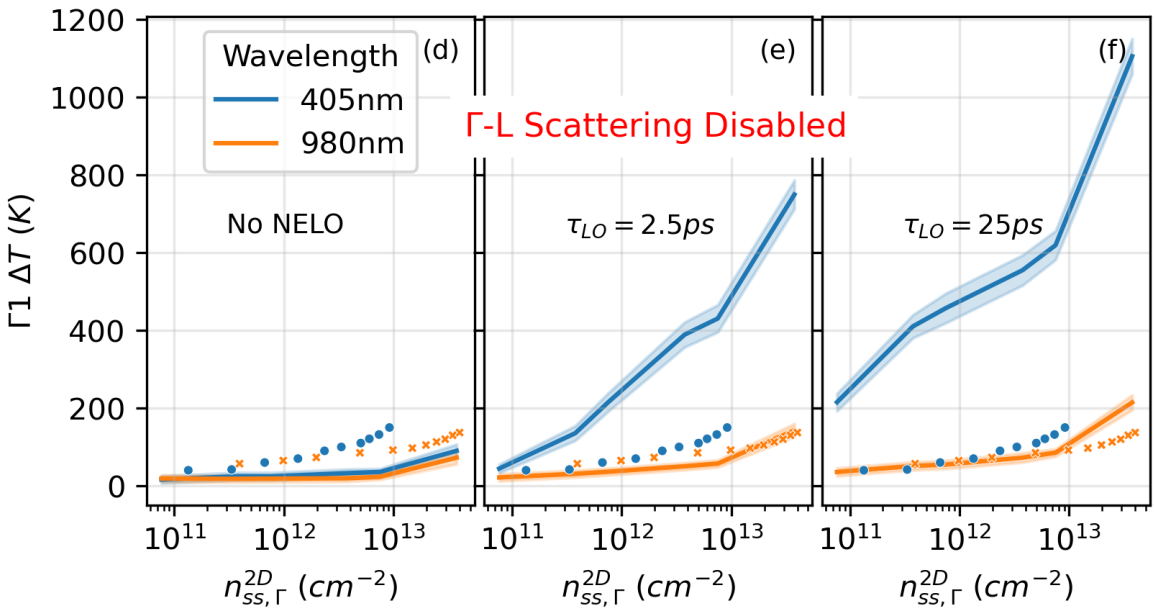

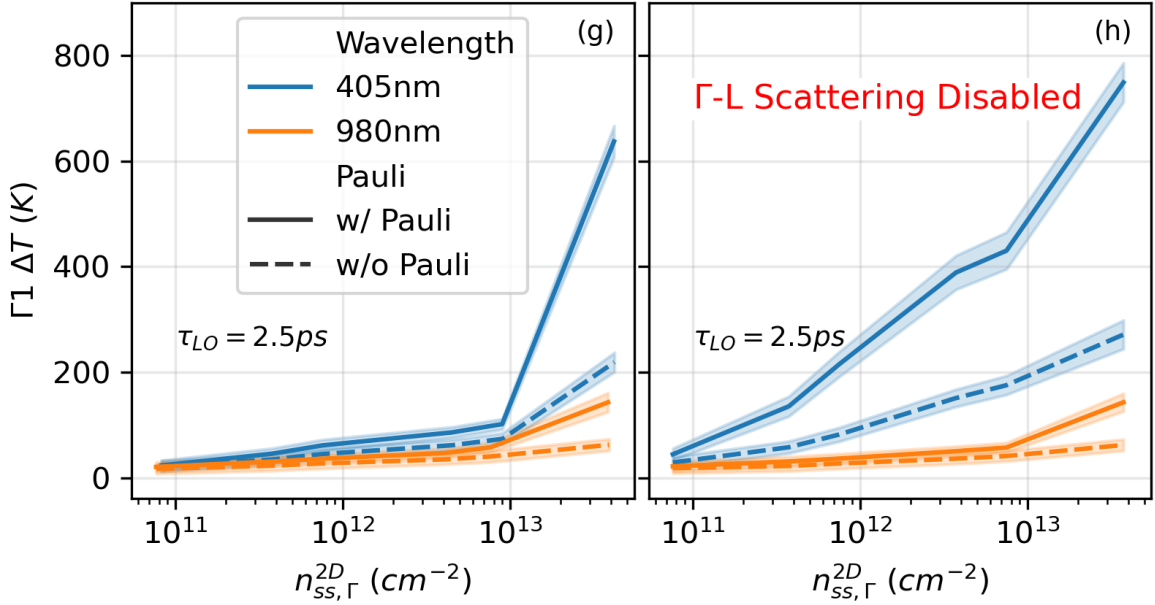

Fig. 6. EMC steady-state temperatures of electrons w.r.t. 300 K in the $\Gamma 1$ subband with 405 nm and 980 nm CW excitation. (a), (b), and (c) Results with $\Gamma$-L scattering enabled. (d), (e), and (f) Results with $\Gamma$-L scattering disabled. From left to right, results are shown for cases without hot LO effects (LO population is always at the equilibrium value), $\tau_{LO} = 2.5ps$, and $\tau_{LO} = 25ps$, respectively. For (a)-(f), the experimental data are also added as symbols for comparison (solid circles and crosses for 405 nm and 980 nm excitation, respectively). (g) and (h) Comparison for with (solid) and without (dashed) Pauli exclusion with $\Gamma$-L scattering enabled and disabled, respectively. Only $\tau_{LO} = 2.5ps$ is chosen for conciseness here. Note that both experiment and simulation have discrete data points, and that the simulated data are plotted as lines without symbols for ease of comparison, and the standard deviations are plotted as shades around the simulated data points.

We already show in Figs. 6(a), 6(b), and 6(c) that without NELO, the electron pathway in which they temporarily visit the L valley and then come back to the $\Gamma$ valley does not directly contribute to a higher $\Gamma$ electron steady-state temperature when Pauli exclusion is considered. Another question we want to address is whether this electron pathway contributes to a higher or lower $\Gamma$ electron steady-state temperature when both the Pauli exclusion and NELO are considered. To answer this question, we perform simulations with $\Gamma$-L scattering (GLS) disabled in comparison to the results above where GLS is enabled. Intervalley scattering involves both optical and acoustic phonons, hence is not completely disassociated with NELO effects. In order to distinguish the effects of intervalley pathways and of NELO, in this work (including the results presented above), we treat the phonons involved in intervalley scattering as separated from the NELO that interact with $\Gamma$ electrons. With this treatment, the electrons still lose or gain energy if they emit or absorb a phonon to undergo GLS. Since the steady-state rates of these two types of GLS are much lower than the dominant scatterings as shown in Fig. 5, the net energy gain by GLS is about zero for $\Gamma$ electrons. Now that with GLS disabled, the steady-state $\Gamma 1$ temperatures are essentially unchanged for the 980 nm simulations as expected, but are increased for the 405 nm simulations, and this effect increases with the LO relaxation time and with $n_{ss,\Gamma}^{2D}$ [see Figs. 6(d), 6(e), and 6(f)]. Without NELO, 405 nm and 980 nm CW excitations lead to similar the steady-state $\Gamma 1$ temperatures. With NELO, all 405 nm excited electrons stay in $\Gamma$ and contribute to more NELO that can be re-absorbed by $\Gamma$ electrons, resulting in significantly higher steady-state $\Gamma 1$ temperatures compared to cases with GLS. Although the $\Gamma$-L-$\Gamma$ pathway temporarily slows the cooling rate of some high energy electrons, at steady state, only a small fraction of electrons goes through this pathway (small GLS rates). This slowing for a small population is obviously outweighed by the loss in NELO that interact with $\Gamma$ electrons. In reality, GLS still contribute to more NELO, but this contribution at steady state is small, also due to the small GLS rates (Fig. 5). It may still be beneficial to have temporary storage at higher energy valleys if we have an efficient way to extract carriers from these valleys, but this is out of the scope of this work.

To check the impact of Pauli exclusion on electron temperature, the case with $\tau_{LO} = 2.5\ ps$ and with and without GLS is chosen and plotted in Figs. 6(g) and 6(h). Without Pauli exclusion, the electrons are allow to fill the low energy states without limit, and show lower temperatures compared to the case with Pauli exclusion. As expected, this effect increases as carrier density increases. Note that temperature extraction from the data without Pauli exclusion use a fitting algorithm for Maxwell-Boltzmann distribution, instead of for Fermi-Dirac distribution featured above.

### G. Nonequilibrium LO phonon population

Figure 7 provides a look into the NELO population for cases with 405 nm and 980 nm CW excitation with and without GLS. The simulations use $\hbar\omega_{LO} = 33 meV$ as the LO phonon energy for In$_{0.53}$Ga$_{0.47}$As, which is a linear approximation from reported values for GaAs and InAs[50]–[53]. Following Bose-Einstein statistics, the equilibrium LO phonon number at 300 K is $N_{LO,300K} = 0.387$. This means without NELO, the number of LO phonons, $N_{LO} = N_{LO,300K} = 0.387$ for all scattered wave vector, q. This result is indeed obtained in simulations without NELO, but is not shown here as it is trivial to plot. With NELO enabled, $N_{LO}$ will be greater than $N_{LO,300K}$ near q=0 (the phonons are hotter than 300 K), as the polar optical phonon scattering rate has an inverse relationship with q, and $N_{LO}$ will decrease and approach $N_{LO,300K}$ as q increases. As shown in Fig. 7, with or without GLS, a 405 nm excitation leads to more NELO than a 980 nm excitation, as higher energy electrons can emit more phonons. For 980 nm cases alone, the NELO populations are essentially the same with and without GLS, since almost no excited electrons have energies high enough to be scattered into the L valley. However, for 405 nm cases, even though the $\Gamma$-L exchange rates are small, the L valley keeps a significant portion of electrons at steady state when GLS is allowed (Fig. 3), leading to much smaller NELO populations compared to when GLS is disallowed. These NELO populations are consistent with the temperature results presented above.

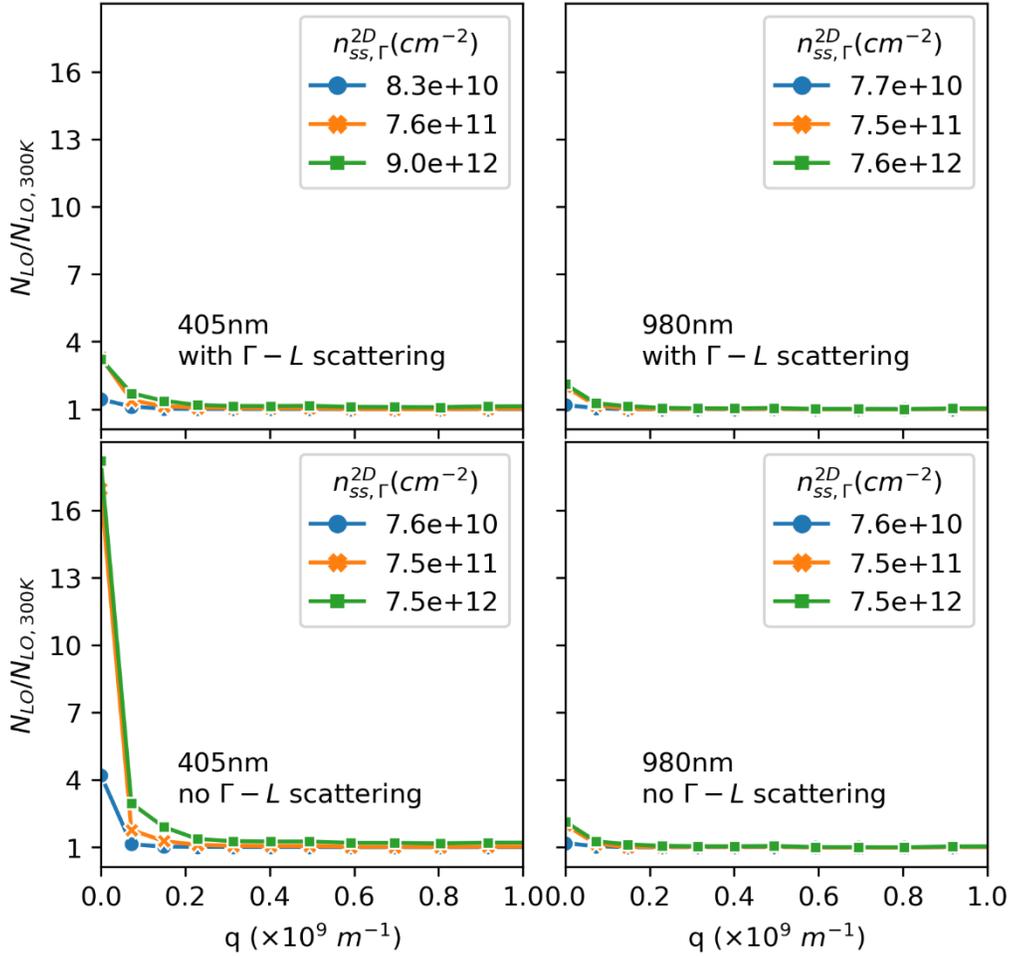

Fig. 7. EMC LO phonon occupation as a function of wavevector q. Results are shown for different excitation wavelengths and carrier densities, with and without intervalley $\Gamma$-L scattering.

## H. Implications

In the search for optimum hot carrier absorbers, one should take into account the energy positions of the valleys. Carrier transfer to higher-energy satellite valleys can reduce the population of nonequilibrium LO phonons, lowering the phonon bottleneck effect and thus carrier temperature in the central valley. For hot carrier solar cells that directly extract current from the central valley, having an absorber that has large energy difference between the central valley and the satellite valleys can minimize the intervalley transfer and maximize carrier temperature in the central valley. On another hand, this work also suggests that increasing carrier temperature in the central valley enhances the carrier transfer from the central valley to the satellite valleys. Therefore, for solar cells that directly extract current from the satellite valleys, the central valley can efficiently give a positive contribution to the current if the valleys are positioned such that $E(satellite) - E(central) kT(central)$. Both scenarios can also benefit from high LO phonon lifetimes in the absorber, in agreement with our prior knowledge of the phonon bottleneck effect. In addition, as shown in Figs. 6(a) and 6(d), the Pauli exclusion effect only has significant impact on carrier temperature at carrier densities higher than $1 \times 10^{13} cm^{-2}$, which corresponds to higher than 10,000 suns and beyond practical solar concentrations. However, photovoltaic LASER power converters [55] may take advantage of this effect to some degree. Finally, at high fermion densities, one can include Fermi-Dirac statistics to ease the analysis for spectroscopic measurements, as there is no requirement for identifying the high-energy tail as with Maxwell-Boltzmann statistics.

## IV. CONCLUSIONS

This work reported the use of the Ensemble Monte Carlo (EMC) simulation method compared with experimental photoluminescence (PL) characterization to investigate the effects of nonequilibrium LO phonons (NELO), Pauli exclusion, and intervalley pathways on the relaxation of $\Gamma$ electrons in an InGaAs MQW for the high-carrier-density regime. The PL results show that 405 nm CW excitation leads to steeper carrier temperature increase as the optical carrier generation rate increases, compared to 980 nm counterparts. The EMC results show the same trend for the steady-state temperature for the $\Gamma 1$ electrons, which can be generalized to all $\Gamma$ electrons. We show that this trend is mainly due to the NELO effect, by comparison of EMC simulations without NELO and with NELO of different decay lifetimes. At electron densities higher than $1 \times 10^{13} cm^{-2}$, the cooling of electron is reduced due to the Pauli exclusion principle, as shown by the scattering rates and steady-state temperature results. Moreover, transfer pathways to a higher-energy satellite valley can reduce the generation of NELO and lead to a lower steady-state temperature of the $\Gamma$ electrons, but also provides another approach to extract hot electrons from the $\Gamma$ valley. This gives us insight for design of optimum hot-carrier absorbers as discussed in III. H.

## ACKNOWLEDGEMENTS

The theoretical work is supported by the National Science Foundation (NSF) and the Department of Energy (DOE) under NSF CA No. EEC-1041895. Any opinions, findings and conclusions or recommendations expressed in this material are those of the author(s) and do not necessarily reflect those of NSF or DOE. The experimental study in France is carried out within the framework of program VI: PROOF (Proof of Concepts of innovative breakthroughs). The authors would like to acknowledge the French ANR project: ICEMAN (No. ANR-19-CE05-0019) for the financial support of the research.

## DATA AVAILABILITY

The data generated and analyzed during this study are available from the corresponding author upon reasonable request.

## AUTHOR CONTRIBUTIONS

**LEGENDS**

Fig. 1. Photoluminescence (PL) characterization of the multi-quantum well (MQW) structure. (a) Schematic of the MQW structure (not to scale). (b) PL spectra of the InGaAs MQW structure at 300 K under various excitation powers of the 405 nm laser. The black lines indicate results of the full spectrum fit. The energy positions of optical transitions in the QW structure are labeled by $E_x$, $E_1$, $E_2$. The $E_x$ is due to excitons. (c) The optical transition energies as a function of absorbed power density. (d) The hot carrier temperature difference $\Delta T$ versus the 2D steady-state carrier density at 300 K under 980 nm and 405 nm excitation wavelengths.

Table 1. Material parameters at 300 K, used for the band structure calculation. They are based on interpolation from Ref.[34], band offset studies[40], and adjustment for quantization effects[35]–[39] according to measured photoluminescent energy peaks in this work. ∥ and ⊥ denote directions parallel and perpendicular to the hetero interfaces, respectively.

Fig. 2. The ensemble Monte Carlo simulation space and mechanisms. (a) The calculated energy levels for the MQW. (b), (c), and (d) show the energy levels (horizontal solid lines) and wavefunctions (dotted curves) of the subbands perpendicular to the layer interfaces for the electrons ($\Gamma$ and L) and heavy holes (HH). (e) The optical generation and recombination, electron-LO phonon interaction, and intervalley scattering processes included in the EMC simulations. Carrier-carrier scattering is also included in the EMC simulations but not illustrated due to drawing difficulty.

Fig. 3. EMC time evolution of electron densities in the $\Gamma$ and L valleys under 405 nm and 980 nm CW excitation. (a) Results without nonequilibrium LO phonons. (b) Results with nonequilibrium LO phonons.

Fig. 4. Electron distribution and temperature fitting. (a) Simulated electron distribution in the $\Gamma$ valley at 1 ns for the lowest ten subbands at $n_{ss,\Gamma}^{2D} = 4.5 \times 10^{12} cm^{-2}$. (b) Simulated electron distribution in the $\Gamma 1$ subband at various $n_{ss,\Gamma}^{2D}$. The arrow in indicates the Fermi energy increases with $n_{ss,\Gamma}^{2D}$. (c) Electron distribution in subband $\Gamma 1$ at 2 ps, 10 ps, and 1000 ps as simulated from EMC (solid cycles) and as Fermi-Dirac fits (lines). The fitted absolute temperatures are indicated in the legend. Failure to fit is default to 0 K. (d) The fitted temperature difference $\Delta T$ as a function of time, with 980 nm excitation and $\tau_{LO} = 2.5 ps$. The colors from blue, orange, green to red, purple, coffee are coded for $n_{ss,\Gamma}^{2D} = 8 \times 10^{10}, 4 \times 10^{11}, 8 \times 10^{11}, 4 \times 10^{12}, 8 \times 10^{12}, 4 \times 10^{13} cm^{-2}$.

Fig. 5. Scattering rates in the $\Gamma$ valley with (left panel) and without (right panel) Pauli exclusion. Both cases have $n_{ss,\Gamma}^{2D} = 4.2 \times 10^{13} cm^{-2}$.

Fig. 6. EMC steady-state temperatures of electrons in $\Gamma 1$ with 405 nm and 980 nm CW excitation. (a), (b), and (c) Results with $\Gamma$-L scattering enabled. (d), (e), and (f) Results with $\Gamma$-L scattering disabled. From left to right, results are shown for cases without hot LO effects (LO population is always at the equilibrium value), $\tau_{LO} = 2.5 ps$, and $\tau_{LO} = 25 ps$, respectively. For (a)-(f), the experimental data are also added as symbols for comparison (solid circles and crosses for 405 nm and 980 nm excitation, respectively). (g) and (h) Comparison for with (solid) and without (dashed) Pauli exclusion with $\Gamma$-L scattering enabled and disabled, respectively. Only $\tau_{LO} = 2.5 ps$ is chosen for conciseness here. Note that both experiment and simulation have discrete data points, and that the simulated data are plotted as lines without symbols for ease of comparison, and the standard deviations are plotted as shades around the simulated data points.

Fig. 7. EMC LO phonon occupation as a function of wavevector q. Results are shown for different excitation wavelengths and carrier densities, with and without intervalley $\Gamma$-L scattering.